\documentclass[twocolumn,aps,xamsmath,amssymb,prd,nofootinbib]{revtex4-1}
\usepackage{graphicx}
\usepackage{epstopdf}
\usepackage{dcolumn}
\usepackage{bm}
\usepackage{hyperref}
\usepackage{color}
\usepackage{amsmath}
\usepackage{url}

\begin{document}

\title{Chiral Dark Sector}

\author{Raymond T. Co}
\author{Keisuke Harigaya}
\author{Yasunori Nomura \vspace{2ex}}

\affiliation{Department of Physics, University of California, Berkeley, California 94720, USA}
\affiliation{Theoretical Physics Group, Lawrence Berkeley National Laboratory, Berkeley, California 94720, USA}

\begin{abstract}
We present a simple and natural dark sector model in which dark matter particles arise as composite states of hidden strong dynamics and their stability is ensured by accidental symmetries.
The model has only a few free parameters. In particular, the gauge symmetry of the model forbids the masses of dark quarks, and the confinement scale of the dynamics provides the unique mass scale of the model.
The gauge group contains an Abelian symmetry $U(1)_D$, which couples the dark and standard model sectors through kinetic mixing. This model, despite its simple structure, has rich and distinctive phenomenology.  In the case where the dark pion becomes massive due to $U(1)_D$ quantum corrections, direct and indirect detection experiments can probe thermal relic dark matter which is generically a mixture of the dark pion and the dark baryon, and the Large Hadron Collider can discover the $U(1)_D$ gauge boson.  Alternatively, if the dark pion stays light due to a specific $U(1)_D$ charge assignment of the dark quarks, then the dark pion constitutes dark radiation. The signal of this radiation is highly correlated with that of dark baryons in dark matter direct detection.
\end{abstract}

\maketitle

{\it Introduction.}---%
Numerous observations consistently point to the existence of dark matter (DM), which constitutes most of the matter content of the Universe. The Weakly Interacting Massive Particle is a highly motivated DM candidate, since the properties necessary to reproduce the observed DM abundance are testable by direct and indirect detection experiments.

In this Letter we propose a model of Weakly Interacting Massive Particles based on chiral gauge symmetry, where all the masses in the model originate from the confinement scale $\Lambda$ of the gauge theory, and the stability of DM is guaranteed by accidental symmetries. The model is remarkably simple, having only a few free parameters.  While Ref.~\cite{Harigaya:2016rwr} considers this model for $\Lambda \sim \mathcal{O}(0.1-1~{\rm GeV})$, we present in this Letter that taking $\Lambda \sim \mathcal{O}(1-100~{\rm TeV})$ leads to rich phenomenology, including indirect detection of two-component DM, direct detection of DM, resonant production of a new gauge boson at colliders, and an observable amount of dark radiation.  A combination of these signals allows us to probe the structure of the dark sector in future experiments.

{\it Model.}---%
The model consists of four Weyl fermions $\Psi_1$, $\Psi_2$, $\bar{\Psi}_1$ and $\bar{\Psi}_2$ charged under $SU(N)$ and $U(1)_D$ gauge groups, which we refer to as dark quarks.  Their gauge charges are given in the two left columns in Table~\ref{tab:charge}.  For $N>2$ and $a \neq 1$, the theory is chiral and the mass terms of dark quarks are forbidden by the gauge symmetry.  The only mass scale of the model is then the dynamical scale of the $SU(N)$ gauge group $\Lambda$.  This makes the model fully natural and relates the masses of all particles in the dark sector with each other.

At the vacuum the dark quarks condense
\begin{equation}
  \left\langle \Psi_1 \bar{\Psi}_1 
    + \Psi_1^\dag \bar{\Psi}_1^\dag \right\rangle 
  = \left\langle \Psi_2 \bar{\Psi}_2 
    + \Psi_2^\dag \bar{\Psi}_2^\dag \right\rangle
  \neq 0 \, ,
\label{eq:condense}
\end{equation}
breaking the approximate $SU(2)_L \times SU(2)_R$ flavor symmetry down to the vector subgroup $SU(2)_V$.  This yields three Nambu-Goldstone bosons.  One of these bosons is eaten by the $U(1)_D$ gauge boson---the dark photon---since the condensation breaks $U(1)_D$.  The mass of the dark photon is given by
\begin{equation}
  m_{A_D} = e_D (1-a) f_{\pi_D} \simeq \frac{\sqrt{N}}{4\pi} e_D (1-a) m_{\rho_D} \, ,
\label{eq:mAD}
\end{equation}
where $e_D$ is the $U(1)_D$ gauge coupling, $f_{\pi_D}$ is the dark pion decay constant, and $m_{\rho_D}\sim \Lambda$ is the mass of the lightest spin-one resonance, which we call the dark rho meson.
\begin{table}[t]
\begin{center}
\begin{tabular}{c|cc|cc}
                &           $SU(N)$ & $U(1)_D$ & $U(1)_P$ & $U(1)_B$ \\ \hline
 $\Psi_1$       &            $\Box$ &      $1$ &      $1$ &      $1$ \\
 $\Psi_2$       &            $\Box$ &     $-1$ &     $-1$ &      $1$ \\
 $\bar{\Psi}_1$ & $\overline{\Box}$ &     $-a$ &     $-1$ &     $-1$ \\
 $\bar{\Psi}_2$ & $\overline{\Box}$ &      $a$ &      $1$ &     $-1$
\end{tabular}
\end{center}
\caption{Charge assignment of the dark quarks under the $SU(N)$ and $U(1)_D$ gauge groups $(N>2)$.  Here, $\Psi_{1,2}$ and $\bar{\Psi}_{1,2}$ are left-handed Weyl fermions, and we take $0 \leq a < 1$ without loss of generality.  The charges under accidental global symmetries $U(1)_P$ and $U(1)_B$ are also shown.}
\label{tab:charge}
\end{table}

The remaining two Nambu-Goldstone bosons form a complex scalar $\phi$, in analogy with the charged pion in QCD.  We call it the dark pion.  The mass of $\phi$ depends on the charge $a$.  For $a\neq 0$, the flavor symmetry of $SU(N)$, $SU(2)_L \times SU(2)_R \times U(1)_B$, is explicitly broken down to $U(1)_D \times U(1)_P \times U(1)_B$.  Here, $U(1)_P$ is a subgroup of $SU(2)_V$, and $U(1)_B$ is the baryon symmetry of the dark sector, whose charges are shown in Table~\ref{tab:charge}.  The dark pion $\phi$ receives quantum corrections to its mass from $U(1)_D$ gauge interactions~\cite{Harigaya:2016rwr,Das:1967it}
\begin{equation}
  m_\phi^2 \simeq \frac{3a \ln 2}{8\pi^2} e_D^2 m_{\rho_D}^2 \, .
\label{eq:m_phi}
\end{equation}
If $a=0$, the flavor symmetry is explicitly broken only down to $U(1)_D \times SU(2)_R \times U(1)_B$, and the dark pion remains massless.  In both cases, $\phi$ couples to the dark photon through gauge interactions
\begin{equation}
  {\cal L} = (D_\mu \phi) (D^\mu \phi)^\dagger \, ,
\label{eq:A_D-pion-1}
\end{equation}
where
\begin{equation}
  D_\mu \phi = \partial_\mu \phi + i e_D (1 + a) A_{D\mu} \phi \, .
\label{eq:A_D-pion-2}
\end{equation}

Other resonances obtain masses of $\mathcal{O}(m_{\rho_D})$.  In particular, the dark baryons have masses
\begin{equation}
  m_B = c \times \frac{N}{2} m_{\rho_D} \, ,
\label{eq:m_B}
\end{equation}
where $c$ is an $\mathcal{O}(1)$ coefficient.  For QCD ($N=3$), $c \simeq 0.8$.

The dark and standard model sectors communicate via the so-called vector portal~\cite{Holdom:1985ag,Goldberg:1986nk,Fayet:2007ua,Pospelov:2007mp,ArkaniHamed:2008qn}.  We introduce kinetic mixing between $U(1)_D$ and the standard model hypercharge, $U(1)_Y$,
\begin{equation}
  {\cal L} = - \frac{1}{4} B_{\mu\nu} B^{\mu\nu} 
    - \frac{1}{4} A_{D\mu\nu} A_D^{\mu\nu} 
    + \frac{1}{2} \frac{\epsilon}{\cos\theta_W} B_{\mu\nu} A_D^{\mu\nu} \, ,
\label{eq:mixing}
\end{equation}
where $B_{\mu\nu}$ and $A_{D\mu\nu}$ are the $U(1)_Y$ and $U(1)_D$ gauge field strengths, and $\theta_W$ is the Weinberg angle.

After the gauge structure is determined by the number of dark colors $N$ and the $U(1)_D$ charge $a$, the model involves only the following three parameters:\ 1.~the dynamical scale $\Lambda$---the overall mass scale of the dark sector, 2.~the $U(1)_D$ interaction strength $e_D$, controlling the hierarchy between dark pion/photon and baryon masses, and 3.~the kinetic mixing parameter $\epsilon$, connecting the dark and standard model sectors. The first two parameters are relevant for the relic abundance and indirect detection, while the third one is important for direct detection and dark radiation signals.  In principle, the phenomenology of the model is predicted by these three parameters.  In practice, since the theory involves strong dynamics, there are a few numerical factors that we can only estimate---$m_B / \Lambda$ and $f_{\pi_D} / \Lambda$---or must treat as a free parameter---the dark baryon annihilation cross section.  To avoid dark baryon overabundance, $\Lambda$ must be smaller than $\mathcal{O}(100~{\rm TeV})$~\cite{Griest:1989wd}.  We require $e_D \leq 1$ to avoid hitting a Landau pole below typical unification scales of $10^{14\mathchar`-17}~{\rm GeV}$.  If the mixing parameter $\epsilon$ is generated via one-loop quantum corrections, $\epsilon = \mathcal{O}(10^{-3})$ is expected.

For models involving strong dynamics and $U(1)_D$ but with massive quarks, see Refs.~\cite{Strassler:2006im,Alves:2009nf,Lee:2015gsa,Hochberg:2015vrg}.  Models involving strong dynamics with vectorlike quarks and a Higgs portal are considered, e.g., in Refs.~\cite{Appelquist:2015yfa,Appelquist:2015zfa}.  Models in which chiral fermion representations are obtained from larger anomaly-free gauge groups are studied in Refs.~\cite{deGouvea:2015pea,Berryman:2016rot}.

{\it Dark Pion as Dark Matter.}---%
The accidental symmetry $U(1)_P$ ensures the stability of the dark pion $\phi$.  For $a \neq 0$, this renders the dark pion a good DM candidate.  This possibility was considered previously in Ref.~\cite{Harigaya:2016rwr} for $m_\phi \lesssim \mathcal{O}(10~{\rm GeV})$, but here we investigate the region $m_\phi \gtrsim \mathcal{O}(10~{\rm GeV})$.  The freezeout process of dark pions depends on the relative value between the dark pion and dark photon masses.

\begin{figure}
\begin{center}
  \includegraphics[width=0.9\linewidth]{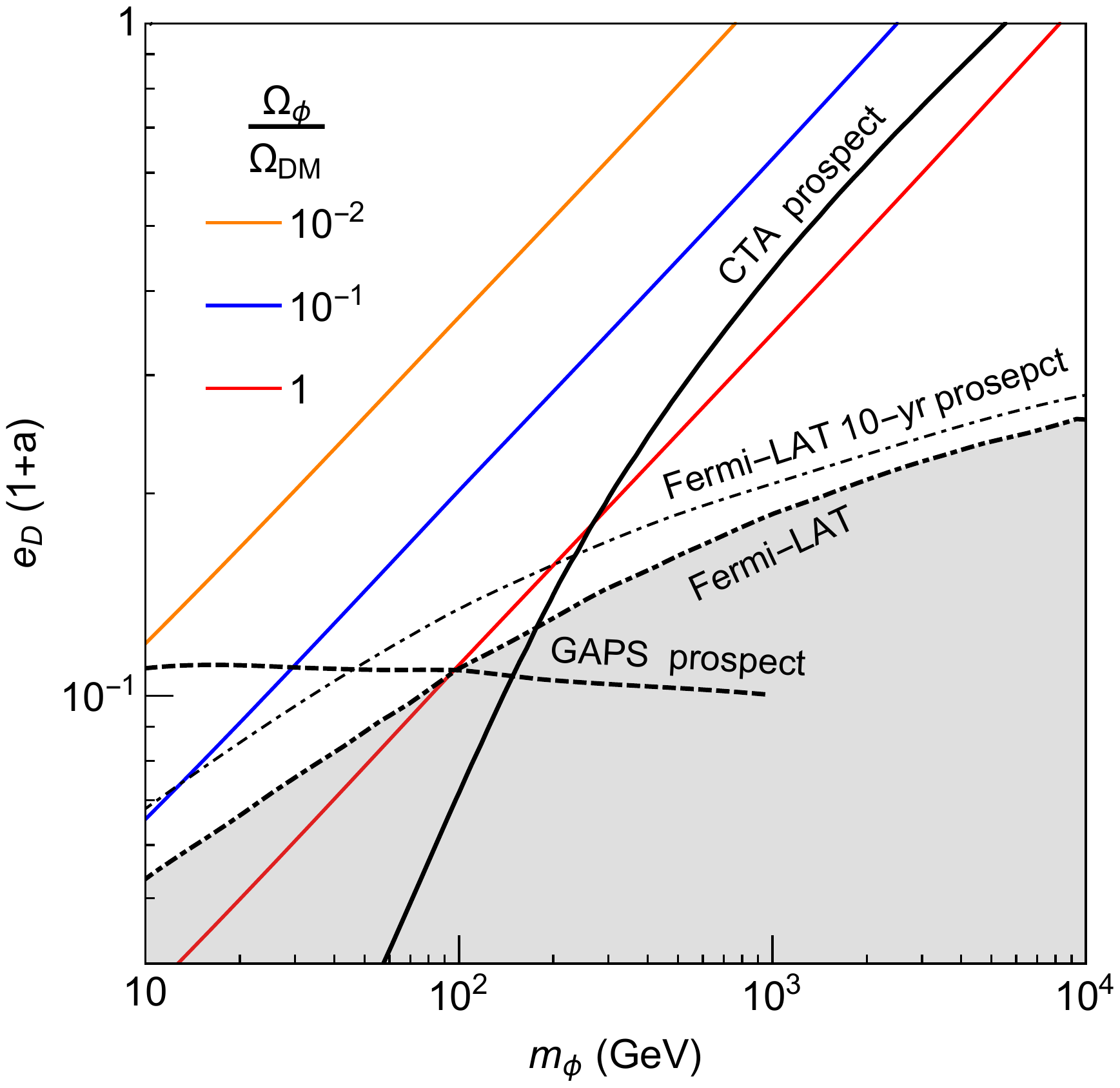}
\end{center}
\caption{Contours of the dark pion abundance $\Omega_\phi$ in units of the value of DM observed today, $\Omega_{\rm DM}$.  Here, we take $f(x) = 1$ for Eq.~(\ref{eq:pion_annihilation}), a very good approximation unless $m_{A_D} \simeq m_\phi$.  As labeled, the black curves show the sensitivities of the current and proposed experiments.}
\label{fig:IndirectSearchesPion}
\end{figure}

In the case where $m_\phi > m_{A_D}$, the dark pion relic abundance is set by the process in which two dark pions annihilate into two dark photons via $U(1)_D$ gauge interactions.  The annihilation cross section is given by
\begin{equation}
  \sigma_\phi v = \frac{3 e_D^4 (1+a)^4}{4\pi m_\phi^2}\ 
    f\biggl(\frac{m_{A_D}^2}{m_\phi^2}\biggr)  \, ,
\label{eq:pion_annihilation}
\end{equation}
where $f(x) = \sqrt{ 1- x} \left( \frac{2}{3} \frac{4-x}{\left(2 - x\right)^2} + \frac{1}{3} \right)$.  As this $s$-wave annihilation mode remains active during big-bang nucleosynthesis and recombination, the dark pion mass less than $\mathcal{O}(10~{\rm GeV})$ is excluded observationally~\cite{Harigaya:2016rwr,Kawasaki:2015peu}.  The annihilation constraint is much weaker for a higher mass (i.e.\ lower number density).  The numerical result of the dark pion abundance is shown in Fig.~\ref{fig:IndirectSearchesPion} as a function of $m_\phi$ and $e_D(1+a)$, which is well approximated by 
\begin{equation}
  \Omega_\phi h^2 \simeq 0.11 \left( \frac{m_\phi}{4~{\rm TeV}} \right)^2 
    \left( \frac{0.7}{e_D(1+a)} \right)^4 \frac{1}{f\bigl(m_{A_D}^2/m_\phi^2\bigr)} \, ,
\end{equation}
where $h \simeq 0.7$ is the Hubble constant in units of $100~{\rm km}/{\rm s}/{\rm Mpc}$.  If we require that dark pions comprise all of the observed DM, $\Omega_\phi = \Omega_{\rm DM}$, then this translates into the prediction of $e_D(1+a)$ in terms of $m_\phi$.  

In the case where $m_\phi < m_{A_D}$, the dark pion annihilates into a pair of standard model fermions $f$ via $s$-channel dark photon exchange, and the cross section is proportional to $\epsilon^2$.  However, the size of $\epsilon$ needed for the correct DM abundance is excluded by direct detection experiments for $m_\phi \gtrsim \mathcal{O}(10~{\rm GeV})$~\cite{Pospelov:2007mp}.  We thus focus on the case with $m_\phi > m_{A_D}$ for $a \neq 0$.  This provides a lower bound on $a$ through Eqs.~(\ref{eq:mAD}) and (\ref{eq:m_phi}).

\begin{figure}
\begin{center}
\includegraphics[width=0.9\linewidth]{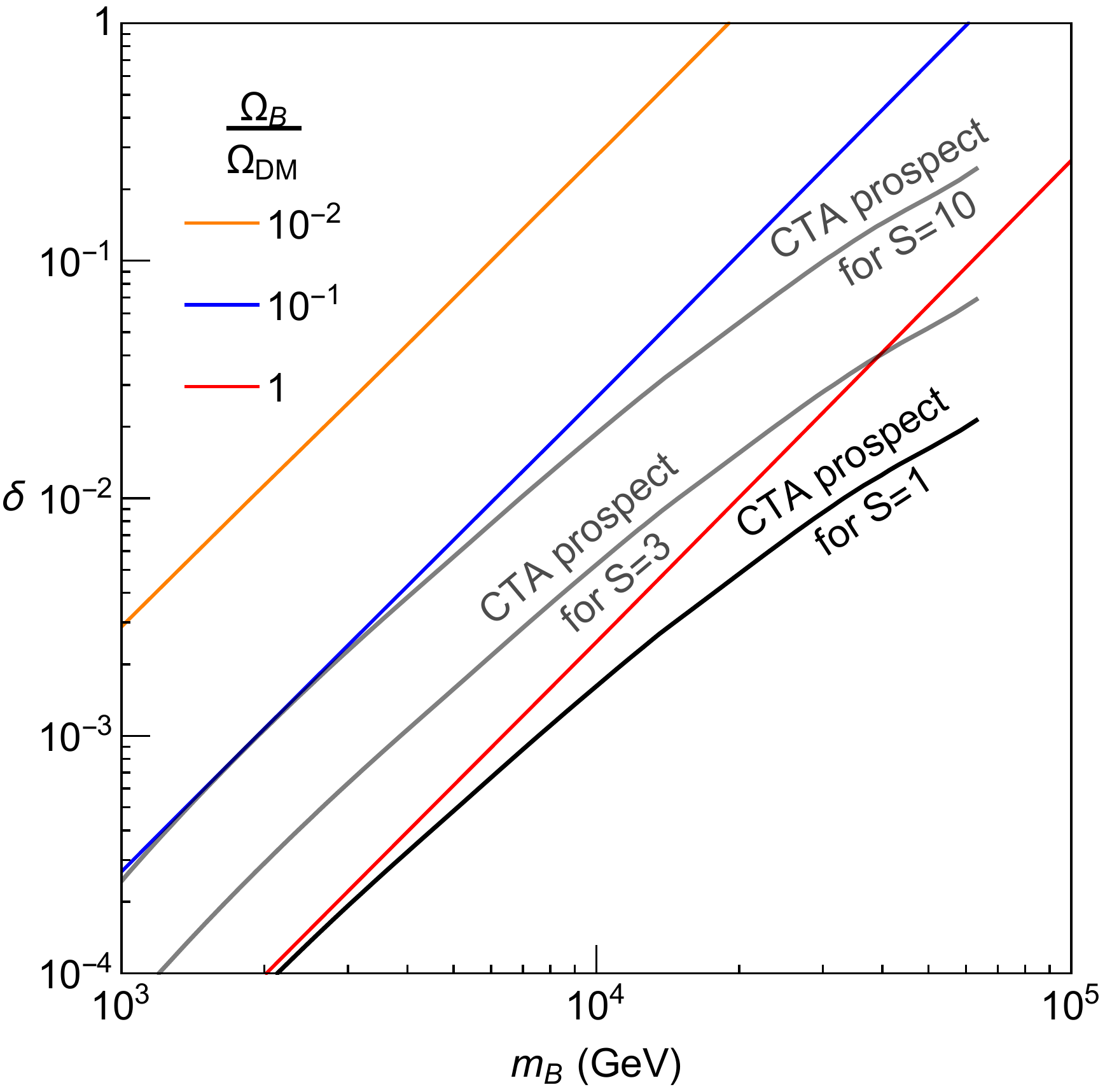}
\end{center}
\caption{Similar to Fig.~\ref{fig:IndirectSearchesPion} but for the dark baryon.  For the projected sensitivity, curves using different values of the Sommerfeld enhancement factor $S$ are shown; see text.}
\label{fig:IndirectSearchesBaryon}
\end{figure}

{\it Dark Baryon as Dark Matter.}---%
The stability of the lightest dark baryon, which we simply call the dark baryon, is guaranteed by the accidental $U(1)_B$ symmetry.  The dark baryon annihilates via dark strong interactions, and its cross section is expected to be large
\begin{equation}
  \sigma_B v = \left(\frac{4\pi}{m_B}\right)^2 \delta \, ,
\label{eq:baryon_sigma_v}
\end{equation}
where $\delta$ is an incalculable factor.  We expect that the dark baryon is well approximated by a point particle at the freezeout temperature, so $\delta$ is subject to the unitarity bound $\delta \lesssim 1/ (4\pi v)$~\cite{Griest:1989wd}.  This factor may be as large as $\mathcal{O}(1)$ around the time of freezeout.  On the other hand, for large $N$ it might be exponentially small, $\delta \sim e^{-\gamma N}$ with some constant $\gamma$, since the annihilation requires one to connect $N$ dark quarks and antiquarks~\cite{Witten:1979kh}. In principle, $\delta$ can be predicted from $N$, e.g.~by lattice calculations, but we treat it as a free parameter.  The numerical result of the dark baryon abundance is shown in Fig.~\ref{fig:IndirectSearchesBaryon} as a function of $m_B$ and $\delta$, which is well approximated by
\begin{equation}
  \Omega_B h^2 \simeq 0.11 \left( \frac{m_B}{20~{\rm TeV}} \right)^2 
    \left( \frac{0.01}{\delta} \right) \, .
\end{equation}

{\it Indirect Detection.}---%
In the present Universe, DM annihilates into dark photons, which decay into standard model fermions via kinetic mixing in Eq.~(\ref{eq:mixing}).  These decay products are observed as cosmic rays.  Here we mainly discuss gamma-ray signals, which are free from uncertainties from propagation.  In our model, DM generally consists of dark pions and dark baryons.

Dark pions annihilate into dark photons with the cross section in Eq.~(\ref{eq:pion_annihilation}).  In Fig.~\ref{fig:IndirectSearchesPion}, we show the current and future sensitivities of the Fermi-LAT satellite~\cite{Ackermann:2015zua,Funk:2013gxa} and the expected sensitivity of the proposed Cherenkov Telescope Array (CTA) observatory~\cite{Carr:2015hta}.  We define the effective cross section
\begin{equation}
  \left(\sigma_\phi v \right)_{\rm{eff}} \equiv \left( \sigma_\phi v \right) 
    \left( \frac{\Omega_\phi}{\Omega_{\rm DM}} \right)^2 \times \frac{1}{2} \, ,
\label{eq:effsigmavpion}
\end{equation}
and compare it with the estimated sensitivities for the annihilation mode ${\rm DM}\, {\rm DM} \rightarrow b \bar{b}$, which is expected to have a similar photon spectrum as our case.  The factor of $1/2$ accounts for the fact that the dark pion is a complex scalar.  The NFW profile~\cite{Navarro:1995iw} is assumed for the CTA sensitivity, which becomes weaker for cored profiles.

Dark baryons annihilate into dark hadrons, which end up as dark pions and dark photons.  As one of the Nambu-Goldstone bosons is eaten by the dark photon, dark photons are efficiently produced by the annihilation of dark baryons.  This is a striking feature of the chiral structure.  We define the effective cross section as
\begin{equation}
  \left(\sigma_B v \right)_{\rm{eff}} \equiv 
    \left( \sigma_B v \right)_{\rm freezeout} 
    \left( \frac{\Omega_B}{\Omega_{\rm DM}} \right)^2 
    \times \frac{1}{2} \times \frac{1}{3} \times S \, ,
\label{eq:effsigmavbaryon}
\end{equation}
where the extra factor of $1/3$ accounts for the fact that dark baryons annihilate both into dark pions and dark photons.  Figure~\ref{fig:IndirectSearchesBaryon} shows the CTA prospect for detecting the annihilation of dark baryons.  From Refs.~\cite{Bedaque:2009ri,Liu:2013vha}, we expect that the dark baryon annihilation receives a Sommerfeld enhancement~\cite{Sommerfeld,Hisano:2002fk} $S$ of order a few due to dark pion exchanges; even $S \gtrsim \mathcal{O}(10)$ may occur if parameters such as $e_D$ are moderately tuned.  We find that the dark baryon can be probed if $S \gtrsim \mathcal{O}(3)$ or if CTA slightly outperforms its expected sensitivity.

\begin{figure}
\begin{center}
  \includegraphics[width=0.9\linewidth]{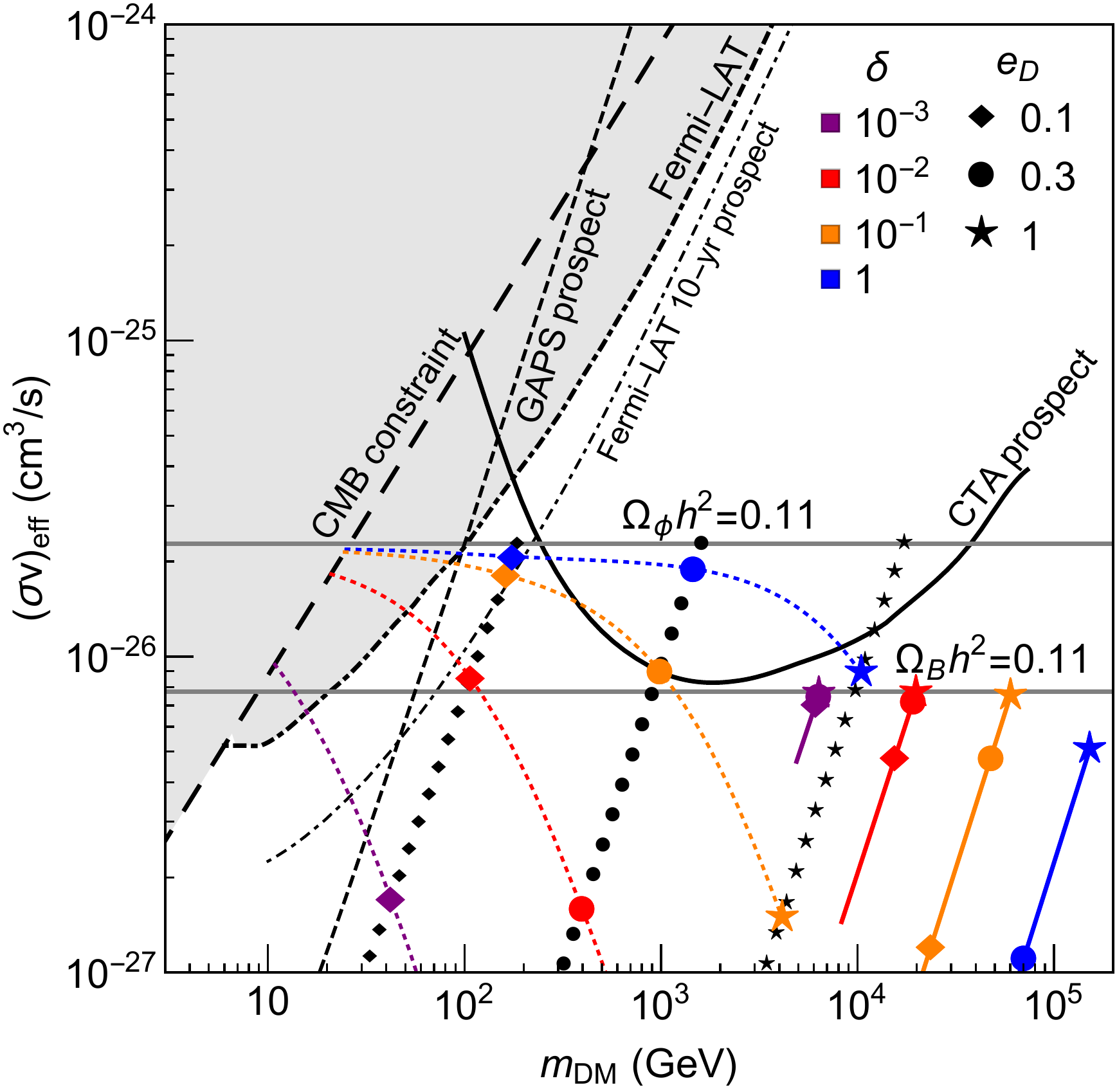}
\end{center}
\caption{The predictions on the masses, $m_{\rm DM}$, and effective annihilation cross sections, $(\sigma v)_{\rm eff}$, of the dark baryon $B$ (solid colored curves) and dark pion $\phi$ (dotted colored curves) for $N=5$ and $a=0.5$, assuming that the DM density today is explained by a mixture of the dark baryon and dark pion.  Each color corresponds to a fixed value of $\delta$, while the symbol shapes indicate the values of $e_D$.  (The black symbols trace out the dark pion properties for fixed $e_D$.)  The two horizontal lines represent the effective cross sections when either the dark baryon or dark pion is the dominant component of DM.  The gray region shows the constraints from the CMB and the Fermi-LAT satellite.  The black curves show the prospects of CTA (solid), GAPS (dashed), and Fermi-LAT (dot-dashed).}
\label{fig:IndirectSearch}
\end{figure}
Figure~\ref{fig:IndirectSearch} shows the correlated predictions on the masses, $m_{\rm DM}$, and effective annihilation cross sections, $(\sigma v)_{\rm eff}$, of the dark baryon (solid colored curves) and the dark pion (dotted colored curves) for various values of $\delta$ labeled by different colors.  Here we require the observed DM abundance to be explained by a mixture of dark baryons and dark pions.  Specifically, for given $\delta$ and $m_B$, we determine $e_D$ and $m_\phi$ from the observed DM density and the mass relations in Eqs.~(\ref{eq:m_phi}) and (\ref{eq:m_B}), using $N=5$ and $a=0.5$ (which is consistent with the condition $m_\phi > m_{A_D}$).  The values of $e_D$ are labeled by symbols along the curves.  The colored curves are truncated on the left (right) by the cosmic microwave background (CMB) constraint~\cite{Kawasaki:2015peu} of $m_\phi \gtrsim 10~{\rm GeV}$ ($e_D \leq 1$).  The dark baryon is the dominant DM component at the locations of the stars, whereas the dark pion dominates toward the other end of each colored curve.  As an example, for $e_D=0.3$ and $\delta = 0.1$, we predict that the dark baryon and pion signals will be discovered at $m_\phi \simeq 1~{\rm TeV}$ and $m_B \simeq 50~{\rm TeV}$ with the corresponding effective cross sections labeled by the orange dots, and that the DM will be a good mixture of both. For different values of $a$ with fixed $\delta$ and $m_B$, the dark baryon prediction remains unaffected, while the predictions associated with the dark pion change as $m_\phi \propto a/(1+a)^2$ and $e_D \propto \sqrt{a}/(1+a)^2$ with $(\sigma v)_{\rm eff}$ unaffected.  These variations are at most of $\mathcal{O}(30\%)$ for $0.3 < a < 1$, so our results are rather robust.

We finally comment on the large antiproton fraction at high energies and the B/C ratio observed by AMS-02~\cite{AMS02antip,AMS02BC}. These data suggest a propagation model leading to a large flux at the Earth~\cite{Giesen:2015ufa}.  If this is the case, the sensitivity of searches with charged cosmic rays may exceed that with gamma-rays.  As an example, we assume the ``MAX'' propagation model and show in Figs.~\ref{fig:IndirectSearchesPion} and~\ref{fig:IndirectSearch} the projected sensitivity of the GAPS experiment~\cite{Fornengo:2013osa}, which goes beyond gamma-ray searches for a low mass region.  In the case where propagation gives a small flux, the large antiproton fraction observed by AMS-02 is well fitted by the annihilation of DM having an $\mathcal{O}(1-10~{\rm TeV})$ mass and a cross section larger than that required for the correct thermal relic abundance~\cite{Ibe:2015tma,Hamaguchi:2015wga,Lin:2015taa} (although it may still be of astrophysical origin, e.g.\ Refs.~\cite{Kachelriess:2015oua,Kohri:2015mga}).  With a modest Sommerfeld enhancement, our dark baryon may be used for this purpose.

\begin{figure}
\begin{center}
\includegraphics[width=0.9\linewidth]{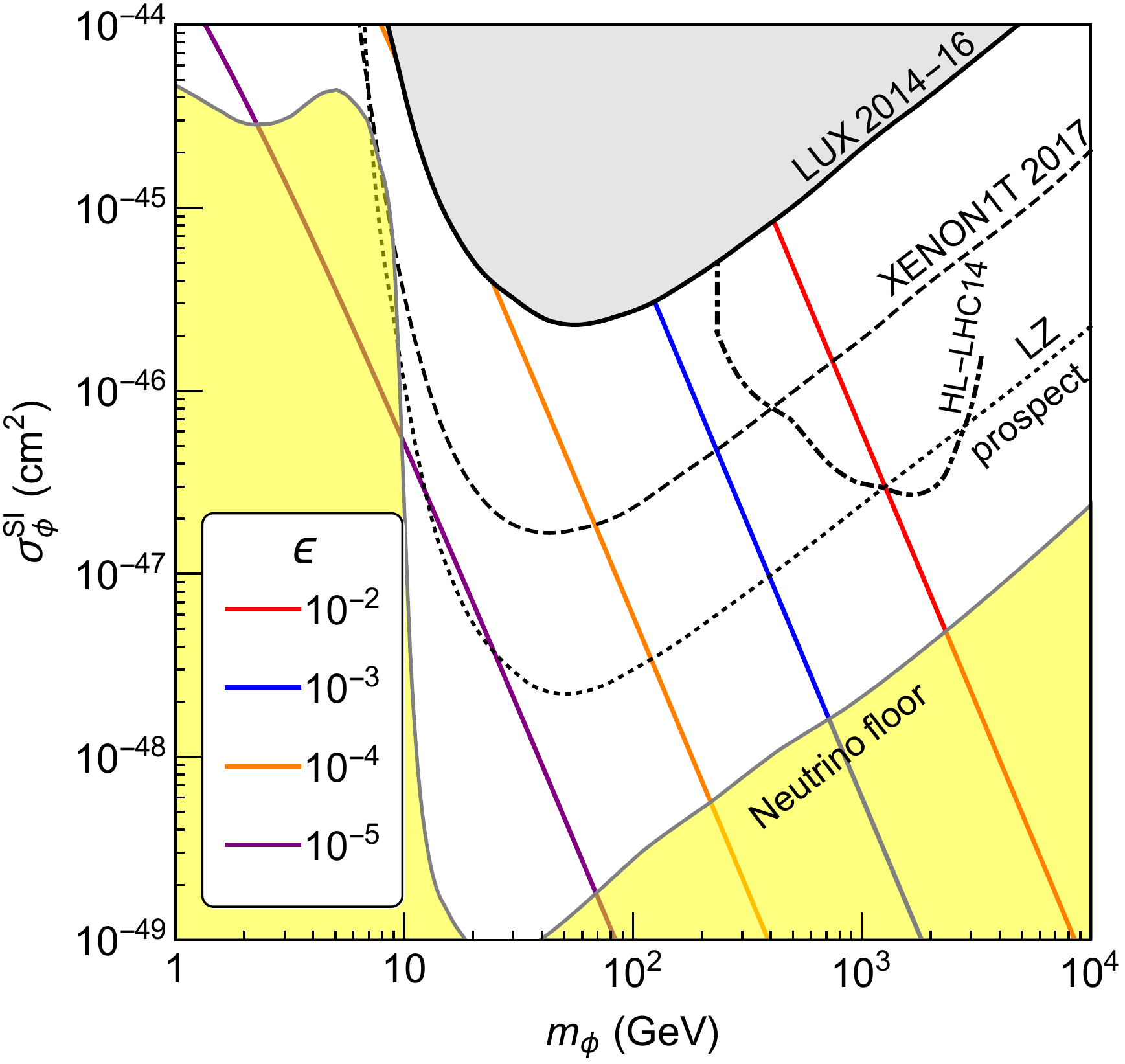}
\end{center}
\caption{The direct detection constraint on the dark pion scattering cross section per nucleon given in Eq.~(\ref{eq:sigmaChiN}).  For each $m_\phi$, we determine $e_D$ such that $\Omega_\phi = \Omega_{\rm DM}$ and $m_{A_D}$ from Eqs.~(\ref{eq:mAD}) and (\ref{eq:m_phi}) using $N=5$ and $a=0.5$.  The dot-dashed line shows the sensitivity of HL-LHC14 to the resonant production of the dark photon, $pp \rightarrow A_D \rightarrow l^+ l^-$.}
\label{fig:DirectSearches}
\end{figure}
{\it Direct Detection.}---%
The dark pion, $\phi$, and dark baryon, $B$, scatter with the standard model protons, $p$, in the nuclei via exchange of virtual dark photons.  The spin-independent dark pion and dark baryon scattering cross sections per nucleon in the non-relativistic limit are
\begin{equation}
  \sigma_{\chi}^{\rm{SI}} = \frac{\epsilon^2 q_\chi^2 e_D^2 e^2 m_\chi^2 m_p^2}{{\pi}(m_\chi + m_p)^2 m_{A_D}^4} \left(\frac{Z}{A}\right)^2 \, ,
\label{eq:sigmaChiN}
\end{equation}
where $\chi = ( \phi, B) $, $q_\phi = (1+a)$, $q_B = (1+a)/2$ (or $0$) for an odd (or even) $N$, and $Z$ and $A$ are the atomic number and atomic weight of the target nucleus.  In Fig.~\ref{fig:DirectSearches}, we compare the dark pion scattering cross section against the current and future direct detection limits~\cite{Akerib:2016vxi,Aprile:2015uzo,Akerib:2015cja,Billard:2013qya}.  If $U(1)_Y$--$U(1)_D$ kinetic mixing is generated via one-loop quantum corrections, $\epsilon = \mathcal{O}(10^{-3})$ is expected.  The dark pion with $m_\phi = \mathcal{O}(100~{\rm GeV})$ can then be detected by near future experiments.

{\it Collider Searches.}---%
Through kinetic mixing, the collision of standard model fermions can resonantly produce the dark photon, which decays back to two fermions.  We show the sensitivity of the high luminosity LHC $14~{\rm TeV}$ run (HL-LHC14, $3000~{\rm fb}^{-1}$) to this process~\cite{Curtin:2014cca} in Fig.~\ref{fig:DirectSearches}.  This provides an interesting correlated prediction between direct detection experiments and collider searches.

\begin{figure}
\begin{center}
\includegraphics[width=0.9\linewidth]{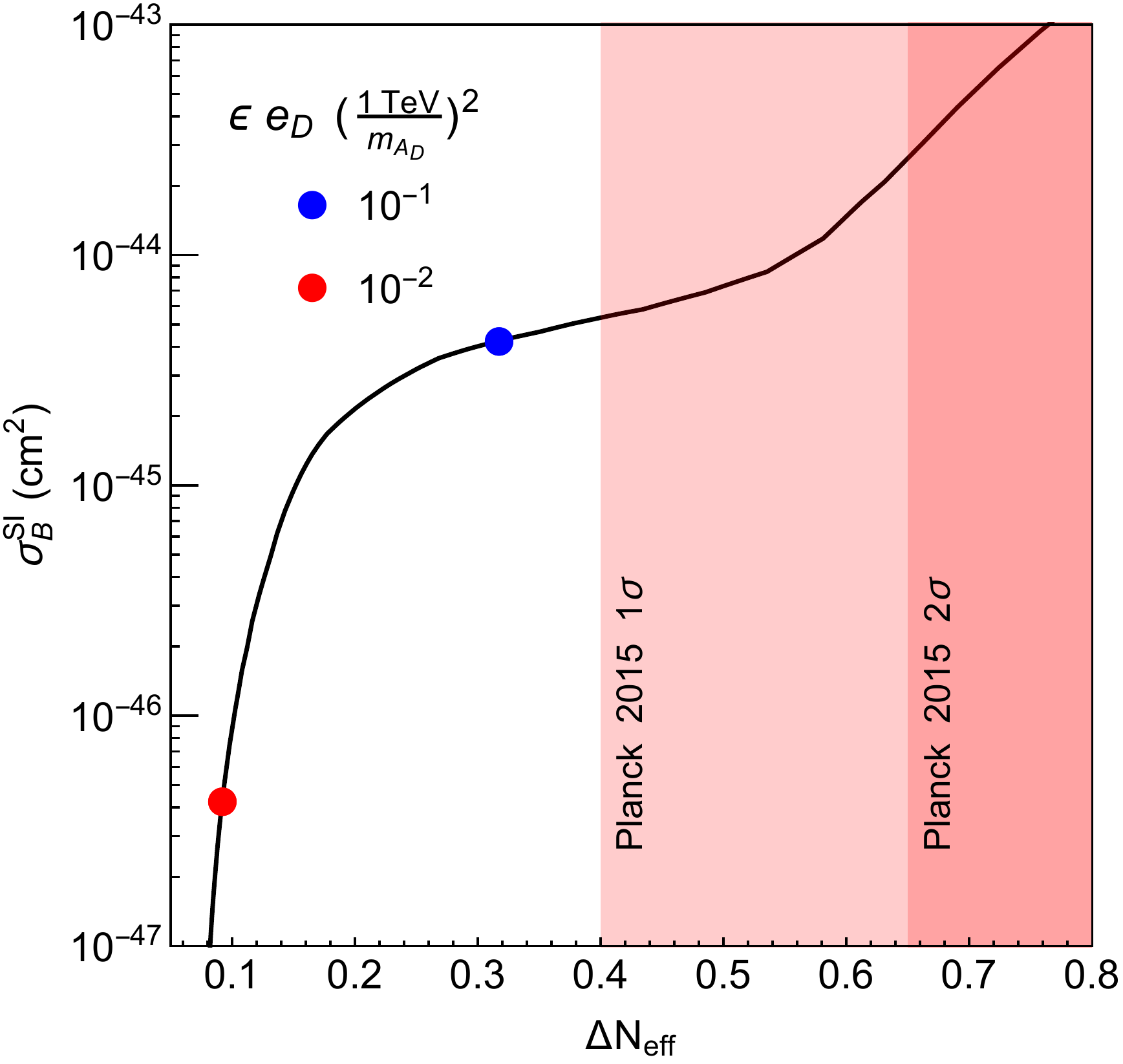}
\end{center}
\caption{The correlated prediction between the direct detection and dark radiation signals for $a=0$ and an odd $N$.  The dots indicate the value of $\epsilon \, e_D / m_{A_D}^2$, which controls both $\sigma_{B}^{\rm{SI}}$ and $\Delta N_{\textrm{eff}}$.  The Planck~2015 constraints~\cite{Ade:2015xua} are shown by the shadings.}
\label{fig:SigmaBNvsDeltaNeff}
\end{figure}
{\it Dark Radiation.}---%
We finally consider the case with $a=0$.  In this case, the dark pion is nearly massless because its mass is no longer generated radiatively by $U(1)_D$ gauge interactions.  This implies that, after decoupling from the thermal bath at temperature $T_{\rm de}$, the dark pion contributes to additional relativistic species
\begin{equation}
  \Delta N_{\textrm{eff}} = 
    \frac{8}{7} \left( \frac{ g_*(1~{\rm MeV}) }{ g_*(T_{\rm de}) } \right)^{4/3} \, ,
\label{eq:DeltaNeff}
\end{equation}
where $g_*$ is the effective entropy degrees of freedom of the standard model thermal plasma, and $g_*(1~{\rm MeV}) = 10.75$ is that around the neutrino decoupling.  The smallest $\Delta N_{\rm eff}$ that Eq.~(\ref{eq:DeltaNeff}) predicts is $0.05$, which can be completely probed by CMB-S4 with the expected sensitivity of $\sigma(N_{\rm eff}) \simeq 0.03$~\cite{Abazajian:2016yjj}.

The dark pion decouples when the scattering rate, $\Gamma = n \sigma v$, drops below the Hubble expansion rate
\begin{equation}
  \frac{16\, \zeta(3) \epsilon^2 e_D^2 e^2}{3 \pi^3 m_{A_D}^4} 
  T_{\rm de}^{5} \simeq \sqrt{\frac{\pi^2}{90}} 
  \frac{\sqrt{g_*(T_{\rm de})} T_{\rm de}^2}{M_{\rm Pl}} \, .
\end{equation}
In estimating $\Gamma$, we have approximated the thermal bath of standard model charged particles as an ideal gas of $e^\pm$, $\mu^\pm$, $\pi^\pm$, and $K^\pm$.  As $\Delta N_{\textrm{eff}}$ and $\sigma_{B}^{\rm{SI}}$ depend on the same combination of parameters $\epsilon \, e_D / m_{A_D}^2$, we can make a distinct correlated prediction between direct detection and dark radiation, presented in Fig.~\ref{fig:SigmaBNvsDeltaNeff}.  Here, the function $g_*(T)$ has been extracted from Ref.~\cite{Borsanyi:2016ksw}.  The value of $\epsilon \, e_D / m_{A_D}^2$ is shown along the curve by the dots.

The mass of the dark pion could be generated by the $SU(2)_R$ breaking dimension-6 operators $\Psi_1\Psi_2 \bar{\Psi}_i\bar{\Psi}_j / M_*^2$, where $M_*$ is the cutoff scale. The resulting mass is
\begin{equation}
  m_\phi \sim \frac{1}{16\pi^2} \frac{\Lambda^2}{M_*} 
  \sim 0.1~{\rm eV} \left( \frac{\Lambda}{100~{\rm TeV}} \right)^2 
    \frac{10^{18}~{\rm GeV}}{M_*} \, .
\end{equation}
The dimension-6 operators with $i=j$ also break the $U(1)_P$ symmetry and hence the two components of the dark pion receive different mass contributions.  These nonzero masses may leave imprints in the CMB and the large scale structure, although the signal largely depends on the size of the dimension-6 operators.

\vspace{0.5cm}
\section*{Acknowledgments}
\vspace{-0.5cm}

We thank Simon Knapen, Surjeet Rajendran, and Sunny Vagnozzi for useful discussions.  This work was supported in part by the Director, Office of Science, Office of High Energy and Nuclear Physics, of the U.S.\ Department of Energy under Contract DE-AC02-05CH11231, by the National Science Foundation under grants PHY-1316783 and PHY-1521446, and by MEXT KAKENHI Grant Number 15H05895.

\end{document}